# Tethered cells in fluid flows – Beyond the Stokes' drag force approach

Johan Zakrisson[†, ‡], Krister Wiklund[†, ‡], Ove Axner[†, ‡], and Magnus Andersson[†, ‡]*

[†]Department of Physics, Umeå University, [‡]Umeå Centre for Microbial Research (UCMR), Umeå University, SE-901 87 Umeå, Sweden;

*Corresponding author: Magnus Andersson, Department of Physics, Umeå University,

SE-901 87 Umeå, Sweden, Tel: + 46 – 90 786 6336, FAX: +46 – 90 786 6673,

E-mail: magnus.andersson@physics.umu.se

## Abstract

Simulations of tethered cells in viscous sub-layers are frequently performed using the Stokes' drag force, but without taking into account contributions from surface corrections, lift forces, buoyancy, the Basset force, the cells' finite inertia, or added mass. In this work, we investigate to which extent such contributions influence, under a variety of hydrodynamic conditions, the force at the anchor point of a tethered cell and the survival probability of a bacterium that is attached to a host by either a slip or a catch bond via a tether with a few different biomechanical properties. We show that a consequence of not including some of these contributions is that the force to which a bond is exposed can be significantly underestimated; in general by ~32-46 %, where the influence of the surface corrections dominate (the parallel and normal correction coefficients contribute with ~5-8 or ~23-26 %, respectively). The Basset force is a major contributor, up to 20 %, for larger cells and shear rates. The lift force and inertia contribute when cells with radii >3 µm reside in shear rates >2000 $s^{-1}$. Buoyancy contributes significantly for cells with radii >3 µm in shear rates <10 $s^{-1}$. Since the lifetime of a bond depends strongly on the force, both the level of approximations and the biomechanical model of the tether significantly affect the survival probability of tethered bacteria. For a cell attached with a FimH-mannose bond and an extendable tether in a shear rate of 3000 $s^{-1}$, neglecting the surface correction coefficients or the Basset force can imply that the survival probability is overestimated by more than an order of magnitude. This work thus shows that in order to quantitatively assess bacterial attachment forces and survival probabilities, both the fluid forces and the tether properties need to be modeled accurately.

## Introduction

Understanding cell adhesion in fluid flows is important in order to elucidate immune cell functions as well as bacterial biofilm formation. However, this is a complicated phenomenon that is strongly regulated by both properties of the bacterial adhesion system and the flow. It involves a multitude of rather complex processes, e.g. receptor-ligand interactions, properties of the mediating structure, and fluid-cell shear forces, of which most are non-linear. For example, the mediating attachment structures often show non-linear elongation-*versus*-force responses while the dissociation rate of the receptor-ligand bonds has a non-linear dependence on the force to which the bonds are exposed, often exponentially within a certain range [1,2]. Moreover, bonds also break stochastically. Since the force adhesion-receptor bonds are exposed to is strongly affected by the flow, this implies that even a small change in the flow can have a large impact on the adhesion lifetime. It has for example been found that blood flow conditions can have a strong influence on leukocyte adhesion to endothelium as well as the regulation of thrombosis of platelets [3].

Unfortunately, it is non-trivial to assess the force or the shear stress an adhering bacterium is exposed to since this is often directly linked to the flow profile, which, in turn, can be both complex and difficult to assess. For example, an increased flow velocity can both provide increased microbial transport towards surfaces and, at the same time, give rise to hydrodynamic forces that can cause attached bacteria to detach [4]. All this implies that the fluid flow is an important factor in microbial deposition.

Although some of the processes that govern bacterial adhesion can be characterized individually by different experimental techniques, e.g., force-measuring optical tweezers and parallel-plate flow chamber instruments [5–7], not all can. This implies that it is far from trivial to experimentally assess correctly and accurately the adhesion lifetime of a bacterium under a given set of conditions. When a complex phenomenon is difficult to assess experimentally, computer simulations based on appropriate models and relevant input parameters are often a powerful complement. However, when such simulations are performed, it is essential they include all phenomena of importance in a correct manner.

However, cell adhesion under hydrodynamic conditions involves complex interactions that can be both difficult to model and time consuming to simulate. To reduce the complexity of the system and the decrease computational time, various types of approximations are often used. Regrettably, the approximations made are not always well justified or appropriate. In particular, force simulations of tethered particles in viscous sub-layers are often uncritically performed by the use of the Stokes' drag force without taking into account surface correction coefficients and force components such as lift, buoyancy, added mass, the Basset force, and the finite inertia of the cell. It is not at all obvious that all these can be neglected when tethered cells attached to a surface are considered; on the contrary, it is most plausible that several of these can modify the force a cell experiences in a given shear flow.

In order to assess the importance of the surface correction coefficients and the various aforementioned, often neglected, force components, henceforth referred to as additional force components, we have in this work developed a numerical methodology for accurate simulations of the force to which an adhesin-receptor bond is exposed. By this, we can assess the individual influence of all these on both the force tethered cells near a surface experiences and the lifetime of the adhesion bond by which they adhere under a variety of conditions.

As is further discussed below, the ranges of system parameters considered were chosen to cover a variety of typical situations. The simulations therefore include a large spectrum of shear rates and cell sizes. The shear rate, $S$, ranges from 10 to 10 000 $s^{-1}$, which encompasses the range of shear rates that has been found in typical physiological flow situations [4]. Cells are often in the 1 – 10 µm range. Those that are commonly exposed to the aforementioned flow conditions often mediate adhesion to a surface via long biopolymers, here alternatively referred to as tethers or organelles, that have their adhesins at the tip and whose lengths are typically in the 1 – 10 µm range [8,9]. The size of the modeled cells and tethers considered were therefore chosen to be in these ranges.

The adhesion organelles can have different mechanical properties, e.g., they can be stiff, elastic, or extendable and the adhesin-receptor bonds by which they bind to host cells can be of different kind, e.g., of slip bond or catch-bond type. The survival probability of adhered cells was therefore investigated for a few well-studied bacterial model systems based on some typical combinations of these types of organelles and bonds. We consider, e.g., tethers with the same properties as those of the helix-like attachment organelle type 1 pili, commonly expressed by uropathogenic *E. coli* (UPEC), as well as those of the non-flexible T4 pili that are expressed by Streptococcus pneumonia [10]. In addition, we used the slip and catch-bond models based upon the PapG-galabiose and the FimH-mannose bonds, respectively [11,12].

# Theory and model

### MODEL SYSTEM

The cell under investigation is modeled as a sphere, constrained by a tether attached to an anchor point on the surface and allowed to move in two dimensions. It is considered to be affected by forces from both the fluid, $\mathbf{F}_F$, and the tether, $\mathbf{F}_T$. As is schematically illustrated in Fig. 1, the flow is modeled as a pipe flow with a viscous laminar sublayer close to the surface with a linearly increasing steady flow profile, given by $\mathbf{v}_f(x,y) = v_{fx}(y)\hat{\mathbf{x}} = Sy\hat{\mathbf{x}}$. We consider cases for which the Reynolds number, Re, defined as $\rho(v_{fi} - v_{bi})2r/\mu$, where $\rho$ is the fluid density, $v_{fi}$ and $v_{bi}$ are the velocities of the flow and the sphere in the direction $i$ (where $i$ is $x$ or $y$), $r$ is the radius of the sphere, and $\mu$ is the dynamic viscosity of fluid, is less than one (more specifically <0.15).

### THE EQUATIONS OF MOTION

Since an object with a finite inertia exposed to a net force will experience acceleration, the equations of motion of the cell (in the $\hat{\mathbf{x}}$ and $\hat{\mathbf{y}}$ directions) can be written as

$$m\frac{\partial v_{bx}}{\partial t} = F_{Fx}(t) - F_{Tx}(t), \qquad (1)$$

$$m\frac{\partial v_{by}}{\partial t} = F_{Fy}(t) - F_{Ty}(t), \qquad (2)$$

where $m$ is the mass of the cell and $F_{Fi}(t)$ and $F_{Ti}(t)$ are the fluid and tether force components, respectively, as illustrated in Fig. 1.

### FORCE COMPONENTS - VALIDITY OF ADDING INDIVIDUAL FORCE COMPONENTS IN THE MODEL

The Navier-Stokes equations (NSE) are a set of time-dependent nonlinear partial differential equations that describe the flow of fluids. To use these equations to simulate cell adhesion or to quantify the force in e.g., cell rolling experiments is time consuming and requires significant knowledge of fluid dynamics. To reduce the complexity in fluidic simulations and fitting to experimental data it is common to use simplified equations that are valid for special cases targeting specific physical conditions. We use in this work an additive approach, i.e., approximate solutions of the NSE, derived for special cases and each representing a specific physical phenomenon, are added into a combined equation, allowing us to study the contribution from different cases separately. Such approach has previously been verified for the low Re numbers that are considered in this work in both computational and experimental works e.g., when describing a submerged sphere colliding with a wall by Yang [13], tethered cells in unsteady fluid conditions by Ryu and Matsudaira [14], lift forces on a sphere in a shear flow near a wall by Krishnan et al. [15], a bubble rising at low Re conditions near a vertical wall by Takemura et al. [16], forces on a rigid sphere near a wall by Zeng et al. [17], and lift forces on bubbles rising at finite Re conditions near a wall by Takemura et al. [18]. More specifically, in the work by Ryu et al, experimental data was compared to both additive models and computational fluid dynamics (CFD) simulations with good agreement. The spatial scales, time-scales, and the force models used in this work are very similar to those. Moreover, Maxey and Riley, showed analytically that the force on a sphere far from any wall in the unsteady case is a summation of the steady Stokes drag, the added mass, and the Basset force [19]. Furthermore, Wakiya showed that the analytical solution of the force on a sphere moving parallell to and near a surface, contains, in an additive manner; a surface corrected Stokes drag force and a wall corrected Basset force [20]. That work was later verified and analytically extended to also include vertical motion [21], which in turn was verified experimentally in the work presented in ref. [22].

The force from the fluid is in the present work assumed to have contributions from the drag force in the proximity of a surface, $\mathbf{F}_D(t)$, a lift force, $\mathbf{F}_L(t)$, a buoyancy force, $\mathbf{F}_B(t)$, the Basset force, $\mathbf{F}_{BF}(t)$, and from added mass, $\mathbf{F}_A(t)$. Due to the discussion above (i.e. for Re numbers close to or below one), we can express the total fluid force as sum of these, i.e., as

$$\mathbf{F}_F(t) = \mathbf{F}_D(t) + \mathbf{F}_L(t) + \mathbf{F}_B(t) + \mathbf{F}_{BF}(t) + \mathbf{F}_A(t), \tag{3}$$

where details of each of the forces are presented below.

As is discussed below, the tether is modeled as either of three types: a stiff organelle; a linear-elastic spring, with a force-elongation response following Eq. (S5) in the supplementary material; or as a helix-like polymer according to the Eqs (S6) – (S8).

**Drag force close to a surface**

The fluid drag on a sphere is often modeled by the Stokes' drag force, given by $6\pi\eta r[v_{fi}(t) - v_{bi}(t)]$, where $\eta$ is the dynamic viscosity. However, although commonly used, this expression is valid only for particles in free flow, i.e., far from any surface. Since the cells considered in this work are tethered close to a surface, the Stokes' drag force has to be modified to account for the presence of the surface. This can be done by so-called near surface correction coefficients. These corrections are, however, developed under steady-flow conditions but are used in this work as an ad hoc approximation, which is supported by the results in ref. [20], to simplify the simulations. Since the corrections are dissimilar in the parallel and the normal directions to the surface, it is suitable to express the corrected drag force, henceforth simply referred to as the drag force, as

$$\mathbf{F}_D(t) = F_{Dx}(t)\hat{\mathbf{x}} + F_{Dy}(t)\hat{\mathbf{y}}, \tag{4}$$

where the two force components are given by

$$F_{Di} = C_i 6\pi\eta r \left[ v_{fi}(t) - v_{bi}(t) \right], \tag{5}$$

where, in turn, $C_i$ is the surface correction coefficient [23].

For a sphere moving parallel to a surface in a shearing flow Goldman et al. [24] presented a correction to the Stokes drag force model valid for large distances from the surface. This model has an error of ~2.5 % for distances ~3 sphere radii and less than 1 % at 7 radii, both relative to the surface. For distances close to surfaces (i.e. for $r/y \approx 1$, where $y$ is the distance from the wall to the center of the sphere) Goldman also developed a lubrication model. Since, in our work, the tethered sphere will be both far and close from the surface during a simulation, we use the interpolation formula derived by Chaoui and Feuillebois [25], which is an extension of the work by Goldman and is valid in both regimes. The correction coefficient is given by

$$C_x[y(t)] = \sum_{i=0}^{N} c_i h^{-i}(t), \qquad (6)$$

where $N$ is the polynomial order and $c_i$ are coefficients that all are given in table S1 in the supplementary materials and where $h$ is the normalized wall distance given by $h = y(t)/r$.

Movements perpendicular to a surface has been investigated by Brenner [26]. Based on the model proposed in that work, Schäffer et al. [27] developed an interpolation formula,

$$C_y[y(t)] = \left\{ 1 - \frac{9}{8}h^{-1} + \frac{1}{2}h^{-3} - \frac{57}{100}h^{-4} + \frac{1}{5}h^{-5} + \frac{7}{200}h^{-11} - \frac{1}{25}h^{-12} \right\}^{-1}, \qquad (7)$$

which for $r/y > 1.1$ deviates less than 0.1% from the result assessed by Brenner [26]. This expression is used in present work.

**Lift force**

Due to shear effects of the fluid and object rotation, a sphere that moves parallel to a surface experiences a force directed perpendicularly to the surface. The lift force produced by a sphere in a shear flow far from any surface was investigated by Saffman [28]. Since in our case the sphere is assumed to be attached to a surface by a tether the influence of rotation on the lift force can be neglected. A model for the lift force on a non-rotating sphere in a shear flow near a wall was later developed by Cherukat and McLaughlin [29]. It is valid when the distance between the wall and center of the sphere, $y(t)$, satisfies the condition $y(t) \ll \min\left(\eta/(\rho_f(v_{fi}(t) - v_{bi}(t))), \sqrt{\eta/(\rho_f S)}\right)$, where $\rho_f$ is the density of the fluid, and $S$ is the shear rate. The lift force for a non-rotating sphere near a wall can thereby be written as:

$$\mathbf{F}_L(t) = F_L(t)\hat{\mathbf{y}} = \rho_f \left[ v_{fi}(t) - v_{bi}(t) \right]^2 r^2 L(t)\hat{\mathbf{y}}, \qquad (8)$$

where $L(t)$ is a solution to the lift integral presented in ref. [29] and is a function that modulates the magnitude of the lift force by linking the time dependent sphere position and velocity to the fluid velocity. The function $L(t)$ is given by:

$$L(t) = A_0(t) - A_1(t)G(t) + A_2(t)G^2(t), \qquad (9)$$

where $A_0(t)$, $A_1(t)$, and $A_2(t)$ are defined by:

$$\begin{aligned} A_0(t) &= 1.7716 + 0.216K(t) - 0.7292K^2(t) + 0.4854K^3(t) \\ A_1(t) &= 3.2397/K(t) + 1.145 + 2.084K(t) - 0.9059K^2(t) \\ A_2(t) &= 2.0069 + 1.0575K(t) - 2.4007K^2(t) + 1.3174K^3(t), \end{aligned} \qquad (10)$$

The coefficients in the functions $A_0(t)$, $A_1(t)$, and $A_2(t)$ are numerically obtained by nonlinear minimization of the lift integral, where the functions $K(t)$ and $G(t)$ are given by:

$$K(t) = \frac{r}{y(t)} = h^{-1}, \tag{11}$$

$$G(t) = \frac{Sr}{\left(v_{fi}(t) - v_{bi}(t)\right)}. \tag{12}$$

**Buoyancy and gravity**

Buoyancy is an upward force exerted by a fluid on an immersed object. The net buoyancy can be expressed in terms of the difference in density between the sphere and the surrounding media, i.e., as

$$\mathbf{F}_B(t) = F_B(t)\hat{\mathbf{y}} = \left(\rho_f - \rho_s\right)Vg\hat{\mathbf{y}}, \tag{13}$$

where $\rho_s$ is the density of the sphere, $V$ is the volume of the sphere, and $g$ is the gravitational constant.

**Basset force**

A body accelerating through a viscous fluid will experience a force known as the Basset force. This is a history force that results from the effect of a lagging boundary layer development due to a change in relative velocity, i.e., acceleration or deceleration of the body moving through a fluid. It is commonly neglected in situations where acceleration is small. However, when high shear flows are present and for initial attachment of cells to surfaces, this term may contribute significantly due to rapid changes of the direction of motion. The Basset force can be expressed as

$$\mathbf{F}_{BF} = -6\pi\eta r^2 \frac{K_H^{3/2}}{\sqrt{\pi\nu}} \int_0^t \frac{d\left[v_{fi}(t) - v_{bi}(t)\right]}{d\tau} \frac{d\tau}{\sqrt{t-\tau}}, \tag{13}$$

where $K_H$ is the near surface correction coefficient, that can be written as

$$K_H = 1 + \frac{1}{h^3}\left[0.375 - \frac{0.3125}{\left(1-2h^2\right)^3}\right] - \frac{3}{\left(1-4h^2\right)^3} - \frac{0.015625}{\left(h-h^3\right)^3} + \frac{3}{\left(1-12h^2+16h^4\right)^3} + \frac{0.375}{\left(3h-16h^3+h^5\right)^3}, \tag{13}$$

ν is the kinematic viscosity, and $t$ is the accumulated time from the start of the simulation [13].

**Added mass**

When an object is accelerated in a fluid medium it has to displace some of the surrounding media and the additional force needed to accomplish this can be interpreted as an added mass to the object. For a sphere the extra force due to the added mass can be modeled as [30]

$$\mathbf{F}_A = -m_A \frac{\partial\left(v_{fi}(t) - v_{bi}(t)\right)}{\partial t}, \tag{13}$$

where $m_A$ is the added mass given by

$$m_A = \frac{2}{3}\pi r^3 \rho_f. \tag{13}$$

Since the functional form of the added mass is the same as that of inertia, given by the left hands sides of the Eqs (1) and (2), it is possible to conveniently account for the added mass by ascribing an effective mass to the sphere, given by the sum of its true and added mass. By this, the effect of inertia (as investigated below) will automatically include also the effect of added mass. In general the contribution of added mass and inertia in the system under study are expected to be small. However, since the motion of a tethered cell is complex, moving with and perpendicular to the flow, we cannot *a priori* disregard their influence on the cell movement and the cell adhesion.

**SIMULATION PROCEDURE**

The cell is initially assumed to be attached to the surface with one of its tethers perpendicularly to the surface, i.e. straight above the tethering point, at a distance from the surface equal to the uncompromised length of the tether (as indicated in Fig. 1). Its initial velocity is set to the flow velocity at the position of the center of the sphere.

As is described in the supplementary material, the equations of motion, Eqs (1) and (2), were solved by applying a finite difference scheme with a secant method to solve for the nonlinear parts of the differential equations. For each time step, first the velocities of the sphere and the force experienced by the tether point were calculated. The Eqs (1) and (2) were then solved numerically by the use of analytical equations describing the velocity in the $\hat{\mathbf{x}}$- and $\hat{\mathbf{y}}$-directions, which are expressed by the Eqs (S3) and (S4) in the supplementary material. The position of the sphere was continuously updated by using the calculated velocities. Finally, the survival probability of the adhesin was derived using the force-time data set in the model described in [11].

# Results

**THE FORCE TO WHICH A TETHERED CELL IS EXPOSED – INFLUENCE OF THE SURFACE CORRECTIONS AND THE INDIVIDUAL FORCE CONTRIBUTIONS**

To investigate the influence of the surface correction coefficients – $C_x$ and $C_y$ – and the additional force components – lift, buoyancy, the Basset force, finite inertia, and added mass – on the force response at the anchoring point of a tether, simulations were for simplicity performed for cells attached by the simplest of attachment organelles, i.e., a stiff linker, for a wide spread of shear rates. For each shear rate, the force response was first simulated using all surface correction coefficients and additional force components, henceforth referred to as a full simulation. To assess the relative importance of each correction coefficient and force component, simulations were thereafter performed with one surface correction coefficient or one additional force component excluded at a time, similar to the approached used by [13]. To be able to compare with the case based on the commonly used Stokes' drag force, a simulation was then also performed in which no surface correction coefficient or additional force component was included. Moreover, to investigate the possible influence of sphere size and tether length on these types of investigations, simulations were made for three combinations of sphere size and tether length.

Figure 2 present results from simulations based on a sphere with a small radius (1 µm) and a short stiff tether (2 µm) for the lowest and highest shear rates considered, i.e., 10 and $10^4$ $s^{-1}$ respectively. The data for two intermediate shear rates, $10^2$ and $10^3$ $s^{-1}$, are presented in Fig. S1 in the supplementary material. The simulations show that, for all shear rates, the force at the anchor point first increases with time before it either levels off or starts to decrease. Moreover, the data also indicate that the full simulations (red dashed curves) resulted in the highest tethering forces, whereas those only based on Stokes' drag force (brown dashed-dotted curves) resulted in the lowest, a few tens of percent lower.

To quantitatively assess the relative importance of the two surface correction coefficients and the additional force components we first evaluate, for each shear rate, the maximum force in the full simulation, which in panel A is at the time marked by the vertical grey line at ~110 ms. An entity referred to as the relative contribution of each surface correction coefficient or force correction term, defined as the relative difference between the simulated force at this particular time for the full simulation and that obtained by the corresponding mode of simulation in which one of the surface correction coefficient or force correction term was excluded, was then calculated. This entity, for the cases considered above and for four different shear rates ranging from 10 to $10^4$ $s^{-1}$, is presented in Table 1.

The table shows, first of all, that the Stokes' drag force expression underestimates the total force by 32 to 38 %. The major contributions to this come from the neglect of the parallel and the normal surface correction coefficients, which separately contributes with either 6-8 % or 25-26 %, respectively, and the Basset force contributing with up to 20% for the highest shear rate. As is expected for the case with a small sphere, none of the other force components contribute significantly to the total force; at the highest shear rates the lift and inertia component contributes with ~0.2 % while at the lowest shear rate the buoyancy provides a contribution of ~0.4 %.

To investigate the influence of sphere size and tether length, simulations were also performed for larger spheres (with 3 and 5 µm radii) and for longer tethers (with 6 and 10 µm lengths). The results of the simulations with a 5 µm sphere equipped with a 10 µm tether for a low and a high shear rate, 10 and $2 \times 10^3$ s$^{-1}$ respectively, are presented in Fig. 3. The reason why not a shear rate of $10^4$ s$^{-1}$ was used as the maximum shear rate in these simulations is that in these cases it provides a situation outside the validity of Eq. 8. The results for two intermediate shear rates, $10^2$ and $10^3$ s$^{-1}$, are presented in Fig. S2. Figure 3 shows that the general form of the response is similar to that of the case with a smaller sphere and a shorter tether. The relative contribution of each surface correction coefficient and each additional force component for this case are presented in Table 2. The table shows that a larger sphere size implies that the contributions from the Basset force, lift force, and buoyancy are larger; for the highest shear rate, $2 \times 10^3$ s$^{-1}$, the Basset force contribute with 19%, whereas the lift force and inertia contributes to the total force to ~1 %. For the lowest shear rate, 10 s$^{-1}$, the buoyancy force contributes with only ~2 % to the total force.

A comparison of the data presented in Table 1 and 2, i.e., for the case with a small sphere and a short tether and that with a larger sphere and a longer tether, reveals that the relative contributions of the surface correction coefficients are similar for the two cases. This thus shows that the relative contributions of the surface correction coefficients are not primarily affected by the properties of the sphere.

The results of the simulations with a 3 µm radius sphere equipped with a 6 µm tether are presented in Fig. S3. The results for this case are similar to that of the small sphere and short tether above.

**THE SURVIVAL PROBABILITY OF A TETHERED BACTERIUM – INFLUENCE OF THE SURFACE CORRECTIONS AND THE INDIVIDUAL FORCE CONTRIBUTIONS**

Uropathogenic *E. coli* have a radius ~0.5-1.0 µm, as is illustrated in the AFM micrograph in Fig. S4. The biomechanical properties of the tethers will modulate the force to which the adhesin-receptor bonds are exposed and thereby their survival probability [23,31]. It has been found that the adhesin-receptor bond of the strain of UPEC that expresses P pili is of slip bond type, while that of type 1 pili is of catch-bond type [11]. Whereas the former type of bonds is assumed to have an exponential dependence on the force to which it is exposed, the latter has the unique property that, under force exposure, the bond can undergo a conformational change and enter a strong and long-lived state, i.e., switch to stationary adhesion or stable rolling at moderately high shear rates. It has been found that both leucocytes and *E. coli* express catch bonds that show shear enhanced binding [1].

Since the adherence of bacteria and the lifetime of the bonds depend strongly and non-linearly both on the type of adhesin bond and on the force to which such a bond is exposed, we found it relevant to assess to which extent the inclusion of the various surface correction coefficients and additional force components discussed above affects the survival probability of tethered bacteria adhering to tissue by both these types of bond. For the slip bond, we used the bond parameters given in Ref. [12]. For the catch bond we used a model developed for the FimH-mannose bond that is expressed on the tip of type 1 pili [11] and previously has been used in tethered bacterial adhesion simulations in viscous sub layers [10].

In addition, since the tethers by which the bacteria adhere can be of several types, e.g., stiff or helix-like, and thereby be modeled in several ways [10,32], it was considered important to also investigate to which extent the choice of tether model affects the calculated survival probability. The influence of tether properties on the survival probability was therefore investigated using all three basic types of tether models; a stiff, an elastic, and an uncoilable helix-like. To account for the increased effective bacteria volume due to dense tether distribution, we assigned in these simulations an effective bacteria radius of 1.5 µm (as illustrated in Fig. S4). The contributions to the drag force from individual tethers are thereby to some extent incorporated in this model.

**Stiff tether**

Based on force-time data obtained from the methodology described above, calculations of the survival probability of a 1.5 µm radius bacterium attached via 3 µm long stiff tether by either a PapG-galabiose slip bond [12] or a FimH-mannose catch bond [11] were performed. The panels A and B in Fig. 4 present results from a simulation with a shear rate of 2300 $s^{-1}$. The data show that for the case with a slip bond (panel A), the survival probability drops rapidly for all simulation cases investigated. However, it can also be concluded that in all cases, except for that when only the Stokes' drag force was used, the survival probability drops to zero before the bacterium reaches the surface (defined as when the center of the sphere comes within $1.2\,r$ from the surface at which the simulation ends). This implies that in this case the simulations predict, irrespectively of the level of approximation, that the bacterium most probably will detach before it reaches the surface. For the case when the bacterium was attached by a catch bond (panel B), on the other hand, the survival probability drops to zero before the cell reaches the surface for the full simulation and the cases when both surface correction coefficients are included (i.e., for the cases: *no lift*, *no buoyancy*, *no Basset*, and *no inertia*), which fully overlap, while, for the cases when the normal surface correction coefficient was excluded, or when only the Stokes' drag force was used, the survival probability remains high (> 0.9). This indicates that the level of approximation, in certain situations, here when the cell is attached by a catch-bond and exposed to a shear rate of 2300 $s^{-1}$, can significantly alter the outcome of a simulation of bacterial survival probability.

Figure 5A displays, as a function of shear rate, the survival probability a bacterium, attached by a stiff tether and a slip bond, at the instance when it reaches the surface, for the cases with a full simulation and Stokes' drag force. The data show that for both simulation models, the survival probability is high for low shear rates (< 1000 $s^{-1}$) and low for high shear rates (> 2000 $s^{-1}$). The former implies that the cell has a large probability to reach the surface before it detaches whereas the latter indicates that the probability that the cell will reach the surface before it detaches is low. However, more importantly, the data also indicate that the survival probability will be significantly overestimated for shear rates in the ~1000 to ~2000 $s^{-1}$ range unless the full simulation is used.

**Extendable tethers – Elastic structure**

In order to investigate the effect of different tether structures on the survival probability of the FimH-mannose catch-bond, the stiff tether was then replaced by a Hookean spring, given by Eq. (S5). Figure 4C displays a simulation performed for an attachment organelle with a stiffness of 200 pN/µm, attached by a catch-bond, for a shear rate of 2300 $s^{-1}$. The data show that the response is similar to the case with a stiff tether (Fig. 4B), i.e., that the full simulation and those in which the normal surface correction coefficient are included predict that the cell should detach before it reaches the surface (i.e. that the survival probabilities, which again overlap, decrease rapidly towards zero before the cell hits the wall). For the cases: *no normal*, and *Stokes'*, the survival probability remains high, indicating that the cell will reach the wall before it detaches. However, it can also be seen (by a comparison of the Figs 4B and C) that for the full simulation or when both surface correction coefficients are included, the survival probability remains high (i.e., close to unity) for a longer time when the tether is elastic than when it is stiff. This is caused by the fact that the elongation properties of the tether initially reduce the force experienced by the adhesin. For the other cases, i.e., with one or both of the surface correction coefficients are excluded the final survival probability is found to be slightly lower for the elastic tether than for the stiff one. The reason for this is that the elongation allows the cell to reside in a flow with higher velocity for a longer time. This implies that the elasticity of the tether is effectively not reducing the load on the adhesin as often is intuitively expected.

Simulations were also made for a ten times higher spring stiffness (2000 pN/µm), as used by Whitfield et al. [33]. The data (not shown) demonstrate that the response of the system is then similar to that from a stiff tether.

**Extendable tethers – Uncoilable structures**
The mechanical properties of helix-like pili have been investigated in some detail by force spectroscopy experiment and a model for its elongation properties has been developed and [34–38]. It has been found that such pili can uncoil at a constant force for several micrometers (seconds) giving rise to ideal damping properties for *E. coli* subjected to fluid flow. The simulations made in this work have been based on the elongation model for type 1 pili that previously was presented by Zakrisson et al. [23]. This implies that the velocity for which the structure uncoils is given by Eq. S7. As is shown by Fig. 4D, even though the time needed to reach the surface is ~2 times longer for a cell expressing a helix-like tether than for any other tethers, it was found that the survival probability stays high (above 99%) for all cases considered. This is caused by the pili uncoiling. Hence, this shows that the unique biomechanical properties of helix-like pili indeed are efficient in damping forces; a bacterium will stay attached irrespectively of which force corrections and force components are included.

However, helix-like pili possess this extraordinary damping ability only as long as the quaternary structure of the pili uncoil; in the cases the pilus becomes fully uncoiled it will act similar to a stiff linker and be affected by the full flow force. Figure 5B illustrates, as a function of shear rate, the survival probability of a cell attached by a catch bond via a type 1 pilus at the instance when the cell reaches the surface, again for the two cases with a full simulation and the Stokes' drag force. For the full simulation (black, solid curve) the pilus becomes fully uncoiled and detaches (most probably) before it reaches the surface for shear rates above ~2700 $s^{-1}$, whereas for the Stokes' drag force (red, dashed curve) it does so only for significantly higher shear rates, above ~5700 $s^{-1}$. This indicates that there is a large range of shear rates, in this particular case from 2700 to 5700 $s^{-1}$, for which the survival probability will be significantly overestimated if only the Stokes' drag force is used.

## Discussion

Force simulations of tethered cells in viscous sub-layers are often performed by the use of the Stokes' drag force but without taking into account contributions from surface correction coefficients, lift force, buoyancy, the Basset force, added mass, or the cells' finite inertia. We investigate here to which extent these correction coefficients and additional force components influence the assessment of the force at the anchor point of a tethered cell and the survival probability of its adhesin-receptor bond mediated of a tether of various types. We show that the consequence of not including some of them can be significant, especially when large cells or high shear rates are considered, and in particular when survival probabilities of tethered bacterial cells are assessed. As a specific but important example, we show how they affect the survival probability of a bacterium attached to a host by a PapG-galabiose slip-bond and a FimH-mannose catch-bond via tethers with various biomechanical properties.

Detailed computational investigations of tethered cells in near wall flow are complex and require a full solution of the Navier-Stokes equations. However, such simulations are extremely computational heavy and therefore approximate methods have been developed. In this work we used an additive approach where the force contributions of various approximate solutions of the Navier-Stokes equations were added to a combined equation. The validity of such approach for the Re numbers considered (<0.15) has been confirmed by several experimental and computational works [13,15–18,22].

A selection of relevant hydrodynamic in vivo conditions were chosen from the literature, e.g. from the urinary tract, arterioles, and arteries as well as data from flow chamber experiments and other computer simulations [33,39–43]. For example, shear rates in urinary flow from the bladder through the urethra can reach 2700 $s^{-1}$ [44] whereas in arterial stenosis shear rates can get as high as 20 000 $s^{-1}$ [45]. In turbulent media, energetic flows can occasionally sweep into the viscous sublayer, producing suddenly increased shear rates for milliseconds-long periods [23,46,47], which agrees with the time scale for the motion of a tethered cell in high shear rates (as is shown in Fig. 4). The maximum shear rates used in this work differed between the various situations considered, but were in all cases below 10 000 $s^{-1}$, limited by the validity of the force models used.

For all shear rates considered, which in no case produce a Reynolds number above unity, this work demonstrates that the conventional expression for the Stokes' drag force can underestimate the force significantly (in the specific cases considered here by up to ~46 %). The discrepancies are primarily dominated by the effects of excluded surface corrections whose parallel and normal coefficients contribute individually by 5-8 % and 23-26 %, respectively, and the Basset force which contributes with up to 20 %. For motion close to a surface the Basset force model used in this work is not modified to take into account surface effects since it is difficult to handle numerically. This slightly overestimates the drag force for objects in proximity of the surface, as discussed in refs. [20] and [21], but do not change the outcome of the analysis. Neglecting the other force contributions does not, in most cases, significantly influence the force to which the adhesion bond is exposed whereby they can be neglected. The lift force and inertia play a role only for the case with a large sphere size and high shear rate. For example, for a cell radius of 3 μm and a shear rate of 5 000 $s^{-1}$, a neglect of the lift force will underestimate the force with ~1 %, and a neglect of the inertia will overestimate the force with ~1%. Buoyancy contribute with ~2% to the total force at low shear rates (≤ 10 s$^{-1}$) and large cells (>3 µm). Note that the Buoyancy force reduces the total force when the motion is in the same direction as the gravitational force. Finally, although there is also a near surface correction to the added mass, this was neglected in these simulations, since it was assumed to be negligible. In conclusions, for the cases when the spheres are small (< 3 µm) or when they are exposed to shear rates below 5000 s$^{-1}$, the effects of lift and inertia can therefore most likely be ignored when adhesion forces are simulated.

However, when survival probabilities of cells attached via tethers are addressed, the simulations show that care must be exercised. The results for a cell expressing a stiff tether and attached by a catch-bond, presented in Fig. 4B, show that the level of approximations can significantly alter the outcome of a simulation of bacterial survival. Neglecting one or both of the surface correction coefficients implies that the survival probability can be hugely overestimated. Similar conclusions can be made when considering an elastic tether (Fig. 4C). Contrary, as is shown in Fig. 4D, when the tether was modeled as a helix-like type 1 pilus, it was found that the survival probability is significantly higher, in fact, it stayed above 99 % independent of whether the surface correction coefficients and additional force components were used or not. The difference between the cases with helix-like and the other two types of tethers can be understood in terms of the fact that the uncoiling properties of the helix-like tether modulate the load on the adhesin [23]. Although both uncoiling lengths and uncoiling forces vary significantly between different types of pili [10,48,49], it can thereby be concluded that uncoiling is an important structural feature of helix-like pili.

Figure 5 presents the survival probability for a bacterium under two specific situations: panel A for a bacterium expressing a stiff tether that adheres by a slip bond; and panel B for a bacterium expressing a helix-like tether that adheres by a catch bond, both as a function of shear rate. The simulations show that, irrespective of type of tether or adhesin, the estimated survival probability is strongly dependent on which simulation model is used. They also demonstrate that a small alteration in shear rate may result in a significant change in the survival probability of a bacterium, especially for those adhering by the helix-like pili. The simulations in this paper thus support previous findings that the survival probability of tether-mediated catch-bonds is strongly affected by the properties of the tether.

A common approach to estimate the drag force of a rolling or tethered cell exposed to a fluid flow in proximity of a surface is to use Goldman's approximation. This approximation predicts that for a cell "touching" the surface, i.e. one for which $y \approx r$, the force should be purely horizontal, given by $F_{G,x} = 1.7 \times 6\pi\eta r^2 S$. This is a fairly accurate assumption when describing spherical cells with tethers that are short (a few nm) in a flow chamber moving along the surface. However, for conditions when the cell is non-spherical or when it is attached via µm long tethers, the use of this approximation will give rise to a significant error in the calculation of the tethering force. For example, for a 2 µm spherical cell tethered with a 1 µm long tether the error is noticeable, >40 %. In addition, the approximation does not take into account the normal drag force that arises from the cell's movement towards the surface, i.e. there is no $F_{G,y}$ component. As shown in this work, neglecting the normal component can significantly underestimate the force to which the anchor point is exposed, in the case considered in this work by 23-26 %. When estimating the drag force of cells close to, but note exactly at the surface, it is therefore important to use the proper approximations. This is especially important when bacteria are modeled since both the shape of such cells can be non-spherical and their tethers can be up to a few µm long.

In conclusion, this work has shown that the simulation models, i.e. the level of approximation, significantly can affect the assessment of adhesion forces and survival probabilities of tethered cells. Lift forces, buoyancy, and inertia (including added mass) play often a minor role in force simulations of tethered cells in flows; lift and inertia force are of importance only in high shear rates ($> 10^4$ s$^{-1}$) while the buoyancy force plays a role in low shear rates ($< 10^2$ s$^{-1}$). On the other hand, our work shows that it is often crucial to include appropriate surface corrections to the Stokes' drag force, the Basset force, and that it is of importance to use appropriate tether properties when tethered particles near a surface are simulated.

## ACKNOWLEDGEMENT

This work was performed within the Umeå Centre for Microbial Research (UCMR) Linnaeus Program supported from Umeå University and the Swedish Research Council (349-2007-8673). It was also supported by the Swedish Research Council (to M.A, 2013-5379) and (to O.A, 621-2008-3280).

## Figures

**Figure 1.** Illustration of a cell tethered to a surface exposed to a fluid flow with a symbolic representation of the nomenclature used in the model. The red point represents both the origin of the coordinate system and the position at which the tether is anchored to a surface. The simulation was terminated at a distance of 1.2 radii above the surface since piliated cells will likely experience multi pili attachment close to the surface.

**Figure 2.** The force experienced by the tether and the anchoring point when a sphere with 1 µm radius is attached to a surface via a 2 µm long stiff tether. The red dashed curve (full simulation) represents a force response that includes all surface correction coefficients and all additional force components, marked *full simulation*. The five subsequent curves, marked *no inertia, no basset, no buoyancy, no lift, no parallel,* and *no normal*, represent simulations performed with one surface correction coefficient or one additional force component excluded at a time, as indicated. The lowest-most curve (*Stokes*) represents a simulation based solely on the uncorrected Stokes' drag force. The panels A and B refer to shear rates of 10 and $10^4$ $s^{-1}$, respectively.

**Figure 3.** The force experienced by the tether and the anchoring point when a sphere with 5 µm radius is attached to a surface via a 10 µm long stiff tether. The various curves represent the same situations as in Fig. 2. The panels A and B refer to shear rates of 10 and 2000 $s^{-1}$, respectively.

**Figure 4.** The survival probability when a 1.5 µm radius cell is attached via a 3 µm long tethers for a shear rate of 2300 $s^{-1}$. Panel A represent the survival probability of the PapG-galabiose slip-bond with a stiff tether. Panels B-D represent the survival probability of the FimH-mannose catch-bond with a stiff tether, elastic tether (Hookean spring), and type 1 pili respectively. In all cases, the curves representing the full simulation and those in which both of the surface correction coefficients are included (i.e., *no inertia*, *no buoyancy*, and *no lift*) overlap.

**Figure 5.** Panel A shows the survival probability of a PapG-galabiose slip bond when a 1.5 µm radius cell is attached via a 3 µm long stiff tether for shear rates of 100 to 2500 $s^{-1}$. The solid black line represents the survival probability for the full simulation, whereas the dashed red curve represents the case where only Stokes' drag force is used. The panel shows that when the Stokes' drag force is used, the survival probability is strongly overestimated for shear rates in the ~1000 to ~2000 $s^{-1}$ range. Under similar conditions, panel B show the case when a type 1 pili and are attached via FimH-mannose catch-bond. Here the survival probability will be strongly overestimated for shear rates in the 2700 to 5500 $s^{-1}$ range.

# Figures

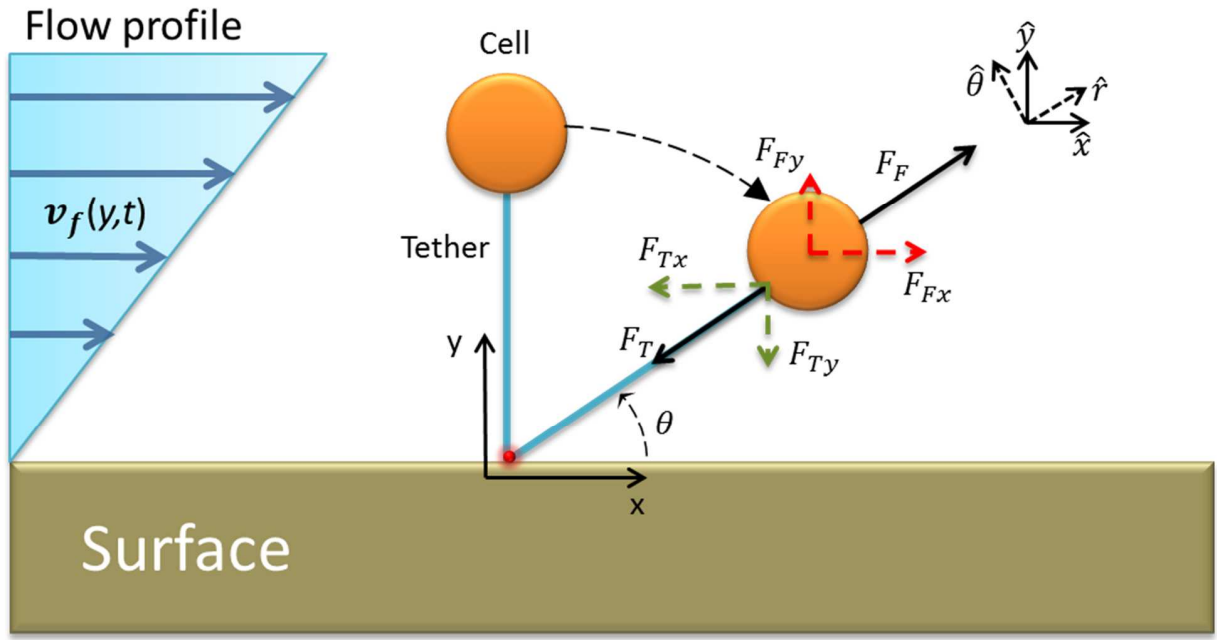

Figure 1

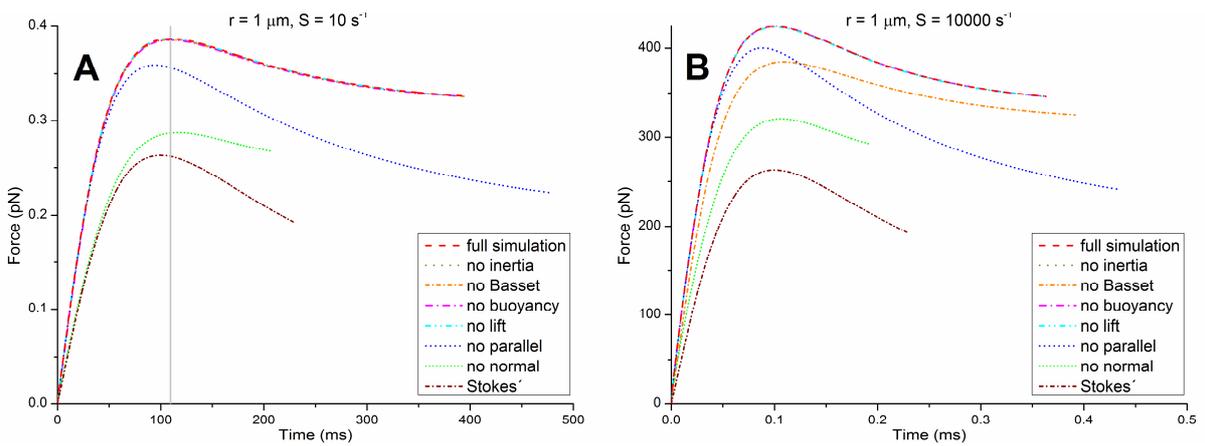

Figure 2

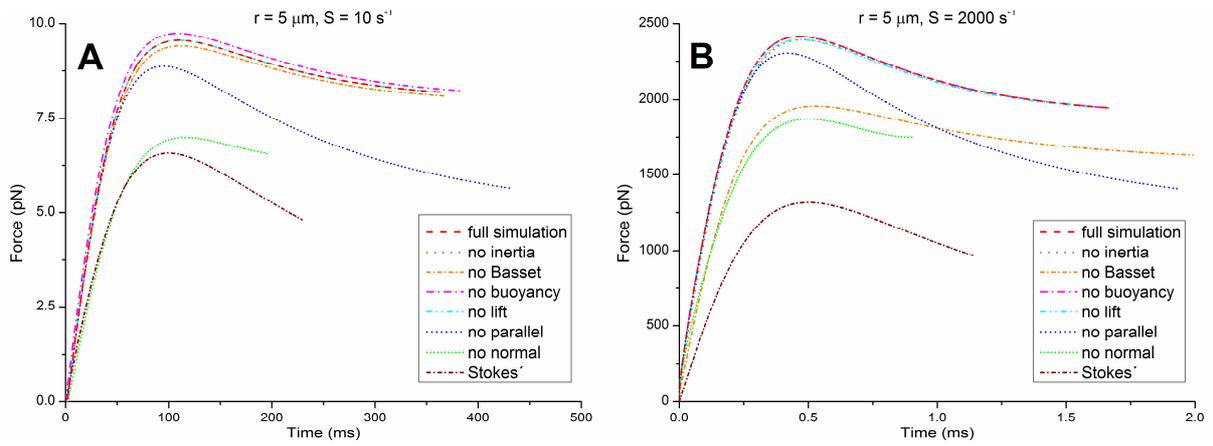

Figure 3

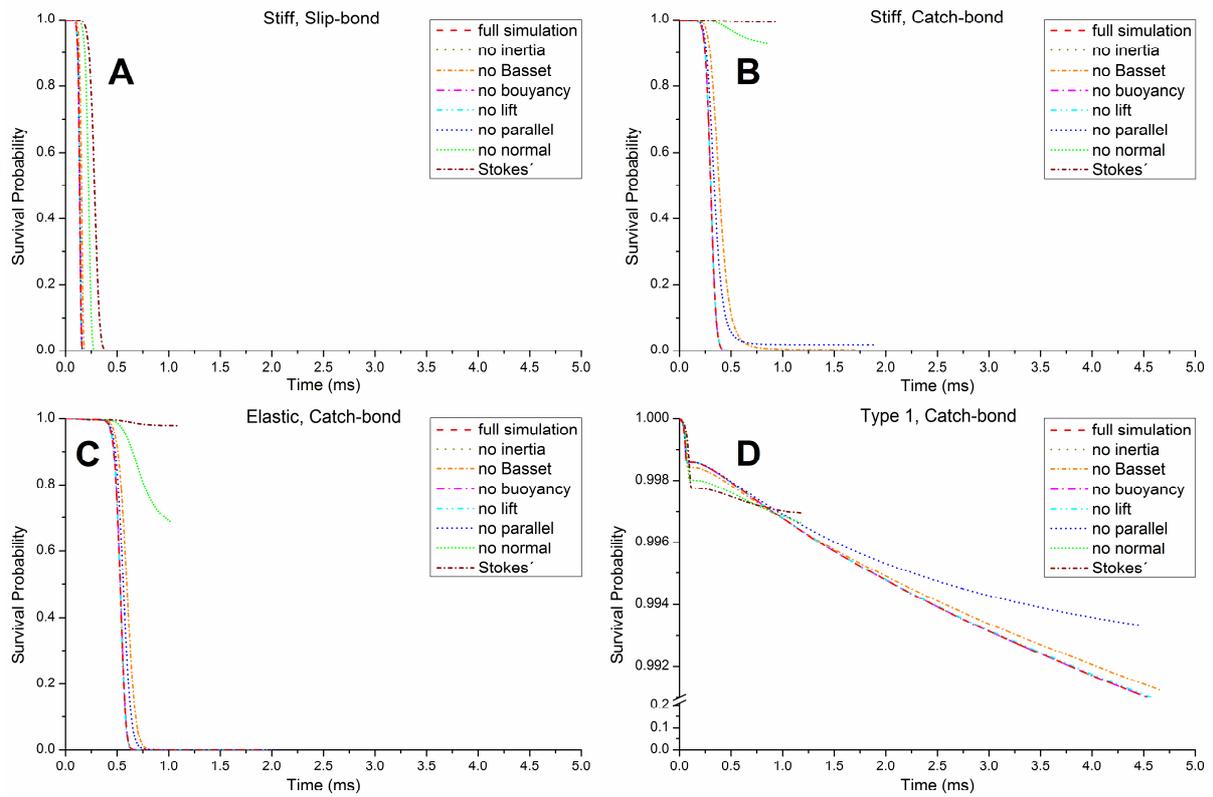

Figure 4

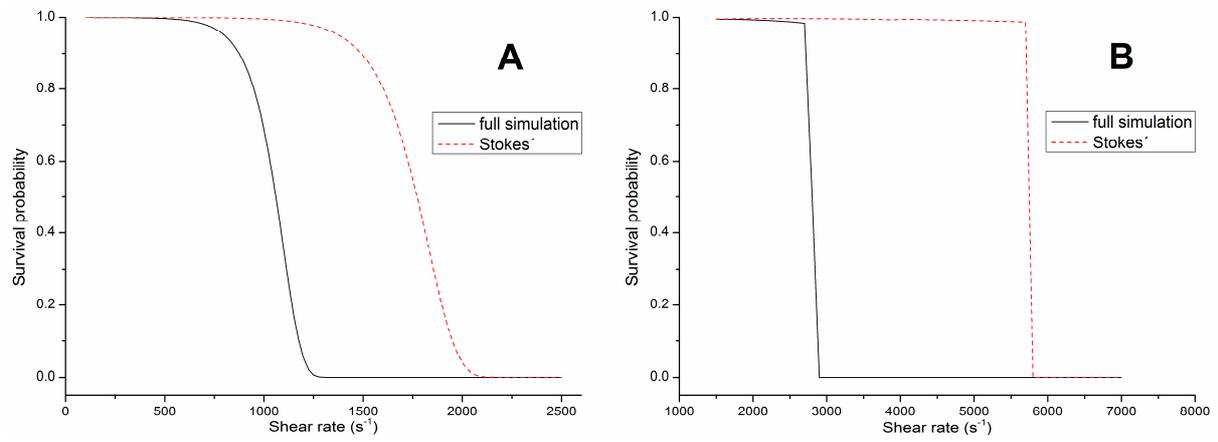

Figure 5

August 11, 2015                                                                                              Manuscript

## Tables

Table 1. The relative force contribution of a given surface correction or additional force component for the case of a sphere radius of 1 µm and a tether length of 2 µm. The relative force contribution is defined as $(F'-F)/F'$, where $F'$ and $F$ are the forces of a full simulation and a simulation with one force component excluded evaluated at the peak force of the full simulation (which for the situation considered in Fig 2A is marked with the dashed line at ~110 ms). The last column represents the relative difference between the force of a full simulation and Stokes' drag force. All results are given in percent.

| Shear rate ($s^{-1}$) | Excluded force component | | | | | | Stokes' drag force |
|---|---|---|---|---|---|---|---|
| | Parallel surface correction | Normal surface correction | Lift | Buoyancy | Basset | Inertia | |
| 10 | 7.9 | 26 | 0.00025 | -0.39 | 0.31 | 0.00013 | 32 |
| 100 | 7.7 | 26 | 0.0025 | -0.039 | 0.97 | 0.0013 | 32 |
| 1000 | 7.4 | 26 | 0.024 | -0.0038 | 3.0 | 0.014 | 34 |
| 10000 | 6.5 | 24 | 0.21 | -0.00035 | 9.3 | 0.17 | 38 |

Table 2. The relative force contribution of a given surface correction or additional force component for the case of a sphere with radius of 5 µm and a tether length of 10 µm. For definition of the relative force contribution, see Table 1. All results are given in percent.

| Shear rate ($s^{-1}$) | Excluded force component | | | | | | Stokes' drag force |
|---|---|---|---|---|---|---|---|
| | Parallel surface correction | Normal surface correction | Lift | Buoyancy | Basset | Inertia | |
| 10 | 8.0 | 27 | 0.0060 | -1.9 | 1.6 | 0.0035 | 32 |
| 100 | 7.2 | 25 | 0.058 | -0.19 | 4.8 | 0.037 | 35 |
| 1000 | 5.8 | 24 | 0.48 | -0.016 | 14 | 0.46 | 42 |
| 2000 | 5.1 | 23 | 0.85 | -0.0075 | 19 | 0.99 | 46 |

23